\documentclass{raa}

\usepackage{graphicx,times}             
\usepackage{latexsym}

\begin{document}

   \title{High Spatial Resolution Spectroscopy of G292.0+1.8 with {\sl Chandra}
 $^*$
\footnotetext{\small $*$ Supported by the National Natural Science Foundation of China.}
}

 \volnopage{ {\bf 2009} Vol.\ {\bf 9} No. {\bf XX}, 000--000}
   \setcounter{page}{1}

   \author{Xue-Juan Yang
      \inst{1}
   \and Xiaoqin Liu
      \inst{1}
   \and Shun-yu Li
      \inst{1}
   \and Fang-Jun Lu
      \inst{2}
   }

   \institute{Faculty of Materials, Optoelectronics and Physics, Xiangtan University,
             Xiangtan 411105, China; {\it xjyang@xtu.edu.cn}\\
        \and
             Particle Astrophysics Center, Institute of High Energy Physics,
Chinese Academy of Sciences, Beijing 100049, China\\
\vs \no
   {\small Received [year] [month] [day]; accepted [year] [month] [day] }
}

   \abstract{
We present high spatial resolution X-ray spectroscopy of supernova
remnant G292.0+1.8 with the {\sl Chandra} observations. The X-ray
emitting region of this remnant was divided into 25 $\times$ 25
pixels with a scale of 20$\arcsec$ $\times$ 20$\arcsec$ each.
Spectra of 324 pixels were created and fitted with an absorbed one
component non-equilibrium ionization model. With the spectral
analysis results we obtained maps of absorbing column density,
temperature, ionization age, and the abundances for O, Ne, Mg, Si, S, and Fe.
The abundances of O, Ne and Mg show
tight correlations between each other in the range of
about two orders of magnitude, suggesting them all
from explosive C/Ne burning. Meanwhile, the abundances of Si and S
are also well correlated, indicating them to be the ashes of
explosive explosive O-burning or incomplete Si-burbing.
The Fe emission lines are not prominent among the whole remnant, and
its abundance are significantly deduced, indicating that the
reverse shock may have not propagated to the Fe-rich ejecta.
Based on relative abundances of O, Ne, Mg, Si and Fe to Si, we
suggest a progenitor mass of $25-30~M_{\odot}$ for this remnant.
   }

   \authorrunning{X.-J. Yang et al.}            
   \titlerunning{{\sl Chandra} observation of G292.0+1.8}  
   \maketitle

%

\section{Introduction}

G292.0+1.8, also known as MSH 11-54, is a bright Galactic supernova remnant (SNR).
It is relatively young with the age of $\le 1600 yr$ (Murdin \& Clark 1979). The detection of strong O
and Ne lines in the optical spectrum (Goss et al. 1979; Murdin \& Clark 1979) classified
G292.0+1.8 as one of the only three known O-rich SNRs in the Galaxy, the others being
Cassiopeia A and Puppis A.

G292.0+1.8 was first detected in X-ray with {\sl HEAO} 1 (Share et al. 1978). With the
increasing sensitivity and resolution of the following X-ray satellites, finer
structures are revealed, such as the central bar-like features (Tuohy, Clark, \& Burton 1982),
metal rich ejecta around the periphery, thin filaments with normal composition centered on
and extending nearly continuously around the outer boundary (Park et al. 2002).
The equivalent width map of O, Ne, Mg, Si clearly show regions with enhanced metallicity of
these elements (Park et al. 2002).
G292.0+1.8 is bright in O, Ne, Mg and Si and weaker in S and Ar with little Fe, suggesting that
the ejecta are strongly stratified by composition and that the reverse shock has not propagated
to the Si/S- or Fe-rich zones (Park et al. 2004; Ghavamian et al. 2012).
Meanwhile, the central barlike belt has normal chemical composition,
suggesting shocked dense circumstellar material for its origin (Park et al. 2004; Ghavamian, Hughes \& Williams 2005).

The detection of the central pulsar (PSR J1124$-$5916) and its
wind nebula (Hughes et al. 2001, 2003; Camilo et al. 2002; Gaensler \& Wallace 2003; Park et al. 2007)
attribute G292.0+1.8 to be a core-collapse SNR. The progenitor mass has been proposed.
Hughes \& Singh (1994) suggested a progenitor mass of $\sim 25~M_{\odot}$ by comparing the derived
element abundances of O, Ne, Mg, Si, S and Ar with those predicted by the numerical nucleosynthesis
calculation based on the {\sl EXOSAT} data. Gonzalez \& Safi-Harb (2003) estimated it to be
$\sim 30-40~M_{\odot}$ with the same method from {\sl Chandra} observation.

In this paper, we present a spatially resolved spectroscopy of G292.0+1.8, using {\sl Chandra}
ACIS observations. In previous work, several typical regions have been picked out and their spectra were
analyzed (e.g. Gonzalez \& Safi-Harb 2003, hearafter GS03; Vink et al. 2004; Lee et al. 2010)
or the equivalent width maps are given (Park et al. 2002).
We, for the first time, give a systematic spectroscopy of this SNR region by region.
In Section 2 \& 3, we present observational data and results. In Section 4, we discuss the nucleosynthesis
and progenitor mass. A summary is given in Section 5.

\section{Observation and Data Reduction}

G292.0+1.8 was observed by ACIS-S3 chip on board {\sl Chandra} on 2000-03-11 from
00:05:46 to 13:21:46 UTC with observation ID 126. After screening the data, the
net exposure time is about 43 ks.

\begin{figure*}
\includegraphics[width=12cm,clip]{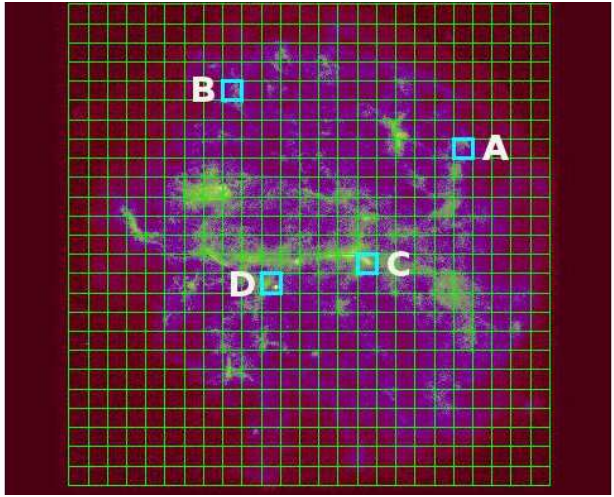}
\caption{
The pixel grid used in our
analyses superimposed on the {\sl Chandra} image of G292.0+1.8.
}
\label{image}
\end{figure*}

The X-ray data were analyzed using the software package CIAO (version 4.1).
In this paper, we divided the X-ray emission region into 25 $\times$ 25
grids with ``pixel'' size of 20$\arcsec$ $\times$ 20$\arcsec$. We extracted events
within energy range 0.5$-$10 keV and created spectra for 328 pixels. Figure~\ref{image} shows
the image of this observation with the grid overlaid. The background spectrum was
created from the off-source region.

The spectra were fitted within XSPEC software package (vesion 11.3, Arnaud 1996),
using one non-equilibrium ionization component (VNEI, Borkowski et al. 2001).
The free parameters are the temperature, emission measure, ionization age ($\tau$ = $n_{e}t$)
and abundances of O, Ne, Mg, Si, S, Fe (in units of solar abundances given by Anders \& Grevesse (1989)).
The reason these elements' abundances being free is that they show prominent emission lines in the
spectra (cf. Figure~\ref{spectra}). We also introduced the WABS model (Morrison \& MacCammon 1983)
for the interstellar photoelectric absorption.

\section{Results}

\begin{figure*}
\includegraphics[width=12cm,clip]{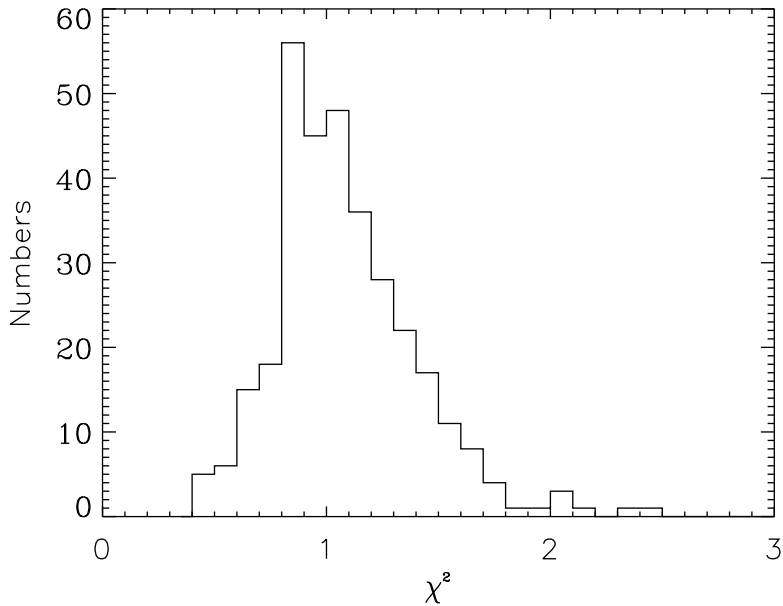}
\caption{
Frequency distribution of the 328 pixel ${\chi}^2$ values
obtained in the spectra fitting.
}
\label{chi_histo}
\end{figure*}

Figure~\ref{chi_histo} shows the frequency distribution of the 328 reduced ${\chi}^2$ values of
the spectral fits, which peaks around 0.85, suggesting the fitting results to be acceptable
statistically.

\begin{figure*}
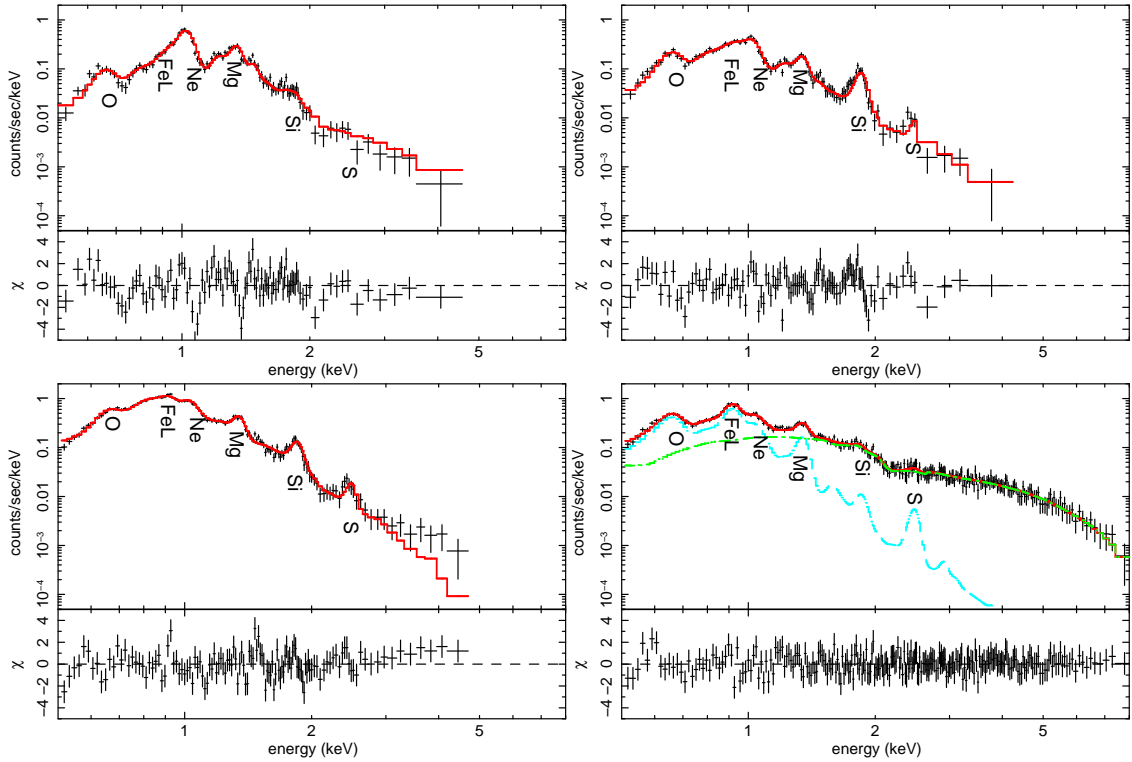

\includegraphics[width=5cm,angle=270,clip]{ms1775fig3_1.ps}
\includegraphics[width=5cm,angle=270,clip]{ms1775fig3_2.ps}
\includegraphics[width=5cm,angle=270,clip]{ms1775fig3_3.ps}
\includegraphics[width=5cm,angle=270,clip]{ms1775fig3_4.ps}
\caption{Spectra along with the fitting results for typical regions as denoted in Figure~\ref{image}. Region A: top left; Region B: top right;
Region C: bottom left; Region D: bottom right. }
\label{spectra}
\end{figure*}

In Figure~\ref{spectra}, we give the spectra along with the fitting residuals of
several typical regions (marked in Figure 1).
Region A is the O/Ne-rich region, with strong O and Ne lines and very weak Si, S lines.
The abundance for O, Ne and Mg are above 4.0 while $\leq$0.3 for Si and S.
As an O-rich SNR, such a region is very typical.
Region B, on the other hand, shows strong Si and S lines, and relatively
weak O and Ne lines. It has a much higher abundance for Si and S ($\sim$ 2.0),
and much smaller abundance for O, Ne and Mg ($\sim$ 3.0) than region A.
Region C is from the central bright belt area, which shows very strong
continuum and medium-mass emission lines. The abundance for all the metal elements
are significantly deduced ($0.3 - 0.5$). Region D covers the central pulsar.
A very strong hard continuum tail is observed from the spectrum.
O, Ne and  Mg lines are clearly shown while no emission lines above 1.8 keV.
Fe-K lines, however, are not detected in any of these regions.

The emission lines from O can been detected in O-rich (A),
Si-rich region (B), and even pulsar-dominate region (D). However, the central
bar-like feature (such as region C), although very bright it is, shows very weak O emission line.
This supports that such a belt should originate from the shocked circumstellar
medium, rather than metal-rich ejecta (Park et al. 2002).
The hard tail in the spectrum of region D is due to the
pulsar. The photon index is 1.87$\pm{0.09}$, which is consistent with the value given by GS03 (1.74$\pm{0.10}$)
and Hughes et al. (2003; 1.6)
and the typical photon index of a pulsar (1.5$-$2.0, Becker 2009).

\begin{figure*}
\includegraphics[width=\textwidth,clip]{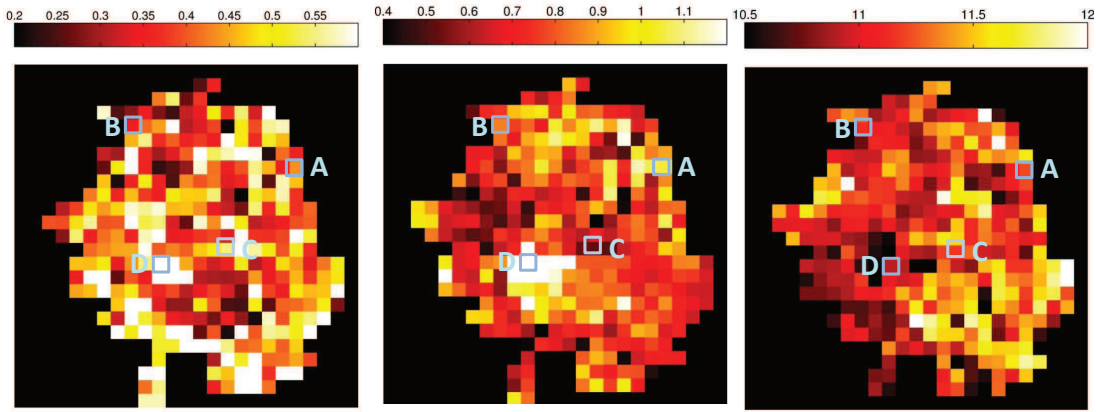}
\caption{The absorbing column density ($N_H$,
$10^{22}$ cm$^{-2}$, left panel), temperature($kT$, $keV$, middle panel)
and ionization age (Log10 $n_{e}t$ cm$^-3$ s,
right panel) maps of G292.0+1.8.
The coding used is shown on the top of each panel.
Regions A/B/C/D as shown in Figure 1 are marked here.}
\label{map_kt}
\end{figure*}

\begin{figure*}
\includegraphics[width=\textwidth,clip]{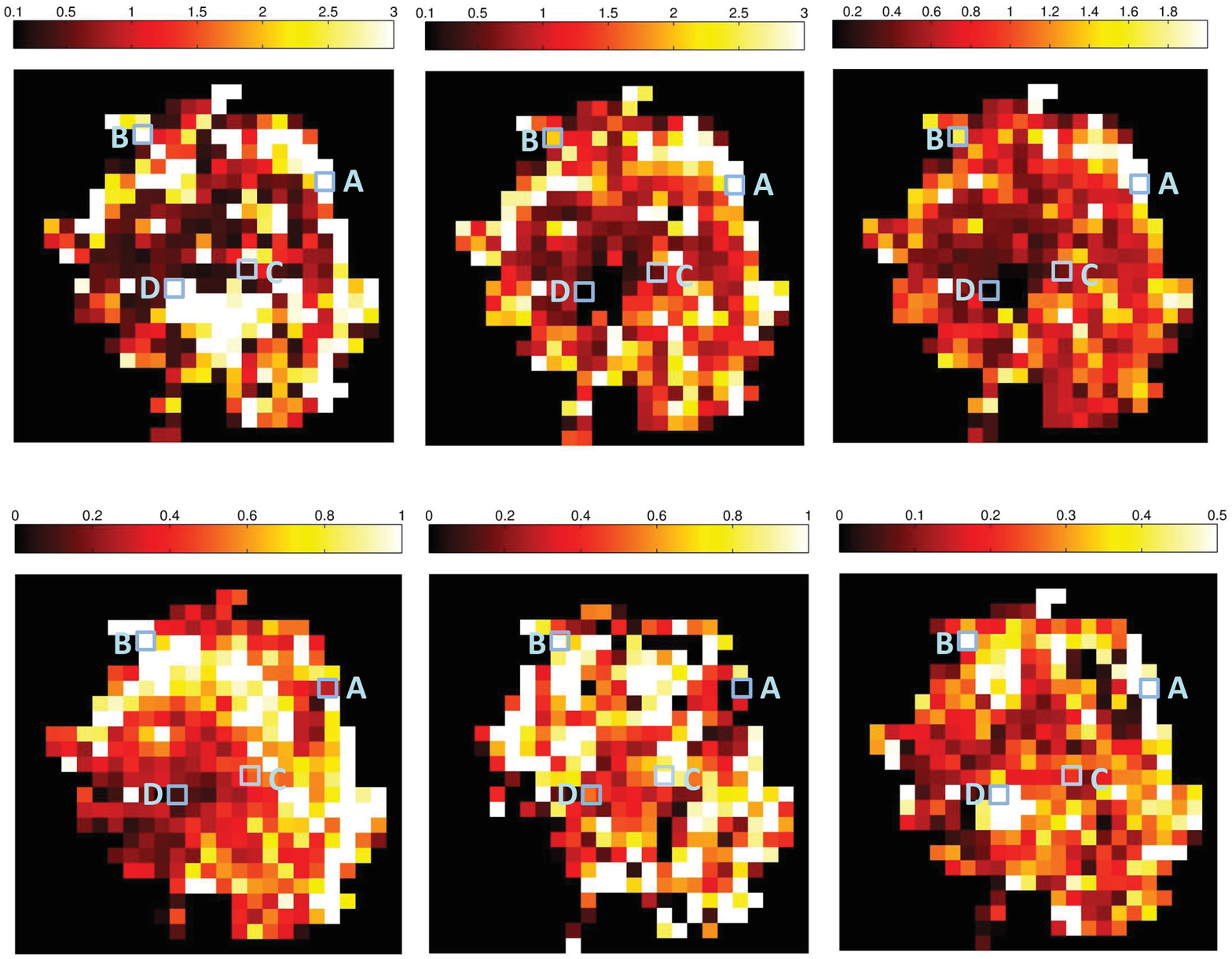}
\caption{O (top left), Ne (top middle), Mg (top right),
Si (bottome left), S(bottom middle), and Fe(bottom right) abundance (in units of solar abundance) maps of G292.0+1.8.
Regions A/B/C/D as shown in Figure 1 are marked here. }
\label{map_abundance}
\end{figure*}

(I) The spatial distribution of the fitted parameters.

The spatial distribution of the absorbing column density, temperature
and the ionization ages are shown in Figure~\ref{map_kt}. The column density
to this SNR lies in the range $0.2-0.7\times10^{22}$ cm$^{-2}$, with the mean
value $\sim~0.45\times10^{22}$ cm$^{-2}$. This is consistent with previous
results (Park et al. 2004).
The column density map is relatively uniform,
suggesting no extra absorbing materials around this SNR.
The temperatures are typically $0.5-1.0~keV$. The relatively ``hot'' regions
are those around the pulsar wind nebula, and the ``high'' temperatures could
be not real but the contamination of the non-thermal emission from the pulsar wind nebula.
The central bright belt of G292.0+1.8 was attributed to the forward
shocked circumstellar medium (Park et al. 2002), which should have larger electron
density ($n_{e}$) and longer shocked time ($t$).
Its ionization age (Log10 $n_{e}t$), however, is even smaller than other region.
The reason might be that these materials have reached ionization equilibrium, and the
ionization age we give here is obtained from non-equilibrium model and thus not
reliable for these regions.

The spatial distribution of all the fitted elements are shown in Figure~\ref{map_abundance} .
It is clear that the central barlike structure is not enhanced in any of these element.
This supports its origin of circumstellar materials (Park et al. 2004; Ghavamian, Hughes \& Williams 2005).
In the meantime, no clear stratification of the element abundances are observed from these maps.
The distribution of the Ne and Si enhanced regions are generally consistent with their EW maps
obtained by Park et al. (2002).

(II) The correlations of the element abundance.

\begin{figure*}
\includegraphics[width=\textwidth,clip]{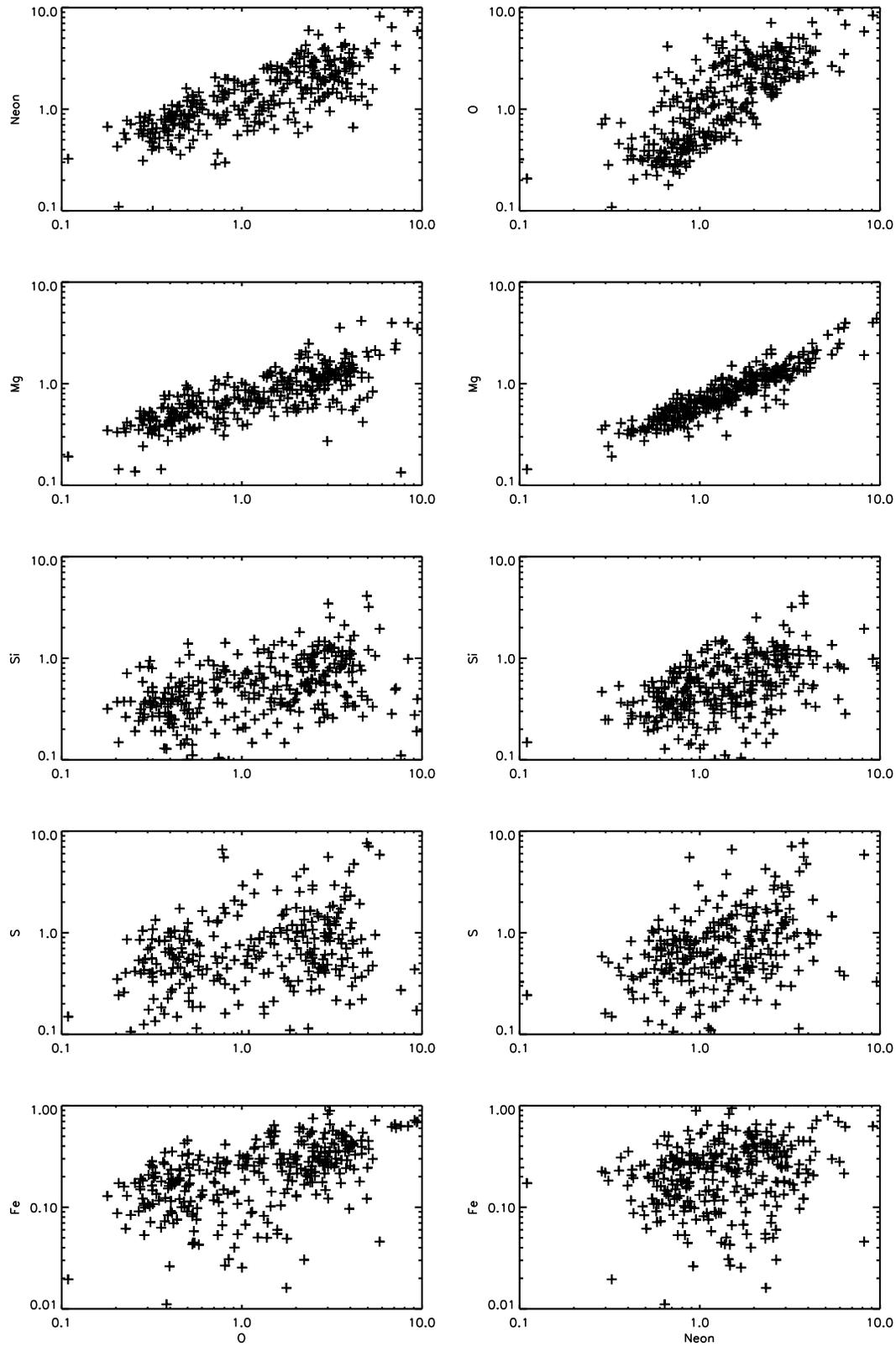}
\caption{Abundance correlation between O/Ne and the other elements.}
\label{Correlate_ONe}
\end{figure*}

\begin{figure*}
\includegraphics[width=\textwidth,clip]{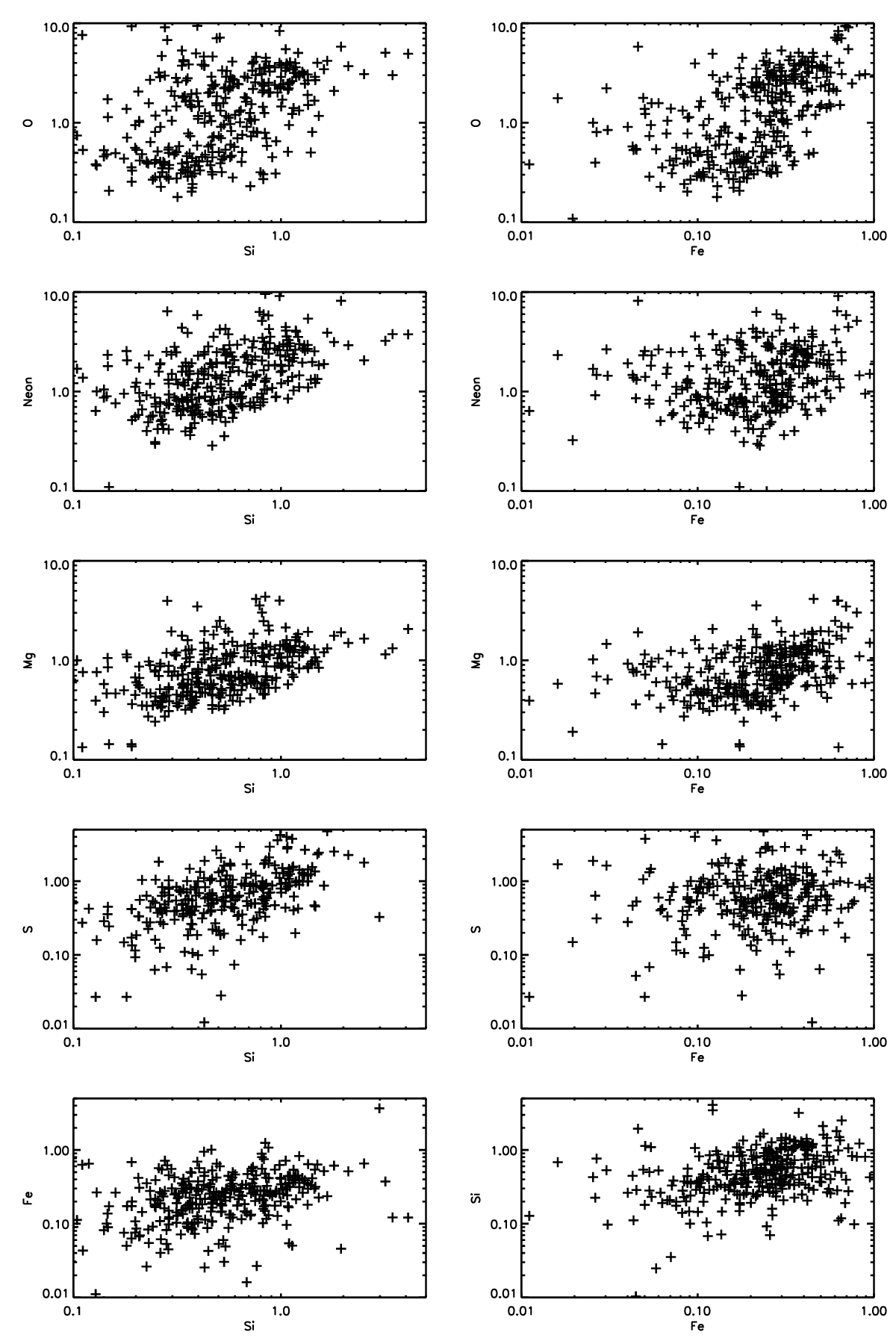}
\caption{Abundance correlation between Si/Fe and the other elements.}
\label{Correlate_SiFe}
\end{figure*}

Figure~\ref{Correlate_ONe} shows the correlation plots of the abundances of
O \& Ne with the other elements, and Figure~\ref{Correlate_SiFe} shows the correlation
of Si \& Fe with the others. The points with minumum element abundances ($\sim 20\%$ of the total data points,
i.e, the very black regions in each of the panel of Figure~\ref{map_abundance})
are excluded, for they represent regions with very weak metal emission lines and thus
probably dominated by the continuum. The coherent coefficients ($\rho$) are given in Table~\ref{coherence_CasA},
along with the corresponding values obtained in Cas A (Yang et al. 2008).

The first observation to Table~\ref{coherence_CasA} and Figure~\ref{Correlate_ONe} $-$ \ref{Correlate_SiFe}
is that in G292.0+1.8 the abundances of O, Ne \&  Mg
are in good correlation with each other, and so is Si {\sl vs.} S. Such a correlation pattern can also
be found in Cas A (cf. Table~\ref{coherence_CasA}), although the coherent coefficients are different.
For example, the most correlated pair in G292.0+1.8 is Ne and Mg, with $\rho = 0.88$, while in Cas A this
coefficient is 0.49. The second observation is that the Fe abundance is basically not correlated with
that of any other element. However, the two phase correlation between Si and Fe abundance found in Cas A
(Figure 8 in Yang et al. 2008) was not shown in G292.0+1.8. These will be discussed in Section 4.1.

\section{Discussion}

\subsection{Nucleosynthesis}

The O, Ne, and Mg abudance show tight correlations with each other within
the range of about two orders of magnitude. This is a strong evidence
that they all come from the explosive C/Ne burning. Meanwhile, such correlation
is also found between Si and S abundance, suggesting them to be ashes
of explosive O-burning and incomplete Si-burning in the core-collapse supernova.
This kind of correlation pattern for elemental abundance is very similar to Cas A (Yang et al. 2008),
and also is consistent with the (explosive) nucleosynthesis
calculations for massive star (Woosley, Heger \& Weaver 2002; Woosley \& Janka
2005). The difference between G292.0+1.8 and Cas A is that the former SNR is
Oxygen-rich while the latter one is Si-rich.

In Cas A, the Fe abundance is positively correlated
with that of Si when Si abundance is lower than 3 solar abundances,
and a negative correlation appears when the Si abundance is higher (Yang et al. 2008).
It is suggested that the Si-rich regions are the ejecta of incomplete explosive
Si-burning mixing with O-burning products, and the regions with
lower Si abundance might be dominated by the shocked circumstellar medium
(CSM). In G292.0+1.8, we can only find a weak positive correlation between
Si and Fe abundance in the whole remnant.
This is not surprising because no prominent Fe lines have been detected
in most of the ``pixels'' we divided for G292.0+1.8, and thus the Fe abundance
can not be that well-constrained. In the meantime, both Si and
Fe are not enriched in the SNR, with their abundances smaller than 1 solar
abundance. This, again, confirms that the reverse shock may have not propagated
into the Fe-rich ejecta (Park et al. 2004; Ghavamian et al.2009; 2012).

\subsection{Progenitor Mass}

Theoretical calculations suggested that for core-collapse supernovae the
different progenitor mass will yield very different abundance pattern (Woosley \& Weaver 1995, etc).
Gonzalez \& Safi-Harb (2003) suggested that the progenitor mass of G292.0+1.8 could be
around $30-40~M_{\odot}$ based on the comparison of the observed abundance ratios with
theoretical values. Their abundance ratios are taken from several ejecta-dominated regions
within this SNR.

\begin{figure*}
\includegraphics[width=\textwidth,clip]{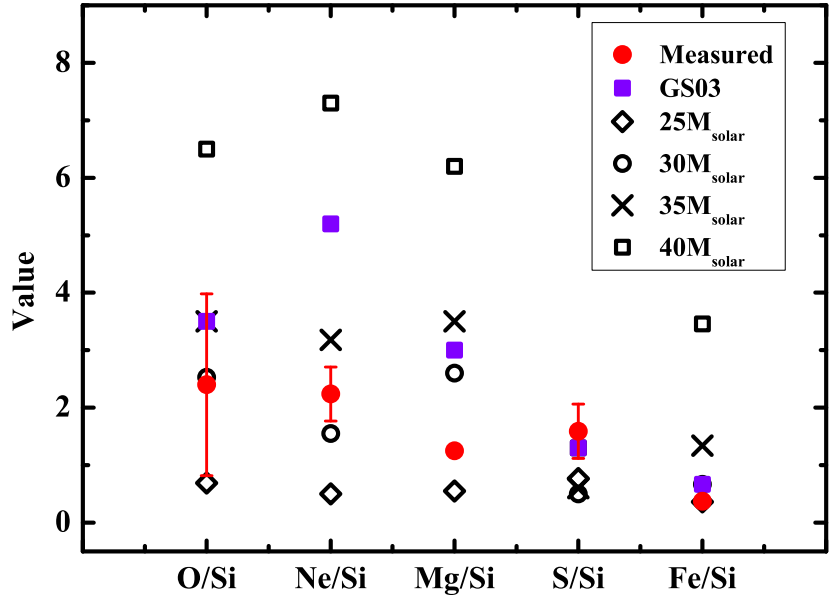}
\caption{Average abundance ratio and their rms scatter,
compared with Gonzalez \& Safi-Harb (2003) and theoretical values.}
\label{G292_progenitor}
\end{figure*}

Here we gave the emission measure weighted average abundance for O, Ne, Mg, Si, S, and Fe among
the whole remnant and their rms ($Z$ and $\sigma_{Z}$ in Table~\ref{abundance_ratio}).
To compare with GS03 work, we also give the abundance ratio of all the elements with respect to
Si and their rms ($Z/Z_{Si}$ and $\sigma_{Z/Z_{Si}}$ in Table~\ref{abundance_ratio}).
By taken G03's observational value and theoretical values they employed, we overplotted
our measurements in Figure~\ref{G292_progenitor}. The abundance ratios measured here
are systematically lower than those from GS03. One would wonder that the reason could be
that the shocked CSM regions are included in our calculation. Considering this, we artificially
choose the ejecta-dominated region by the standard of the O abundance larger than 1.0, and calculated
the corresponding ratio again (Table~\ref{abundance_ratio}, $(Z/Z_{Si})^{\prime}$). We can see
that $(Z/Z_{Si})^{\prime}$ are generally consistent with $Z/Z_{Si}$ within the scatter.
This is not surprising for two reasons. One is that the Si abundance are tentatively correlated
with other elements. The other is that the larger element abundance values contribute much
more to the final average values, so that even though the average abundance are much larger, their ratios
tend to be similar. Our measurements suggested the progenitor mass for G292.0+1.8 to be $25-30~M_{\odot}$,
smaller than $30-40~M_{\odot}$ suggested by GS03.

\section{Summary}

We did a spatially resolved X-ray spectroscopy of the SNR G292.0+1.8.
The spatial distribution of the absorption column density, temperature,
ionization age, and the abundances of O, Ne, Mg, Si, S and Fe are given.
The uniform column density map suggests no extra absorbing material around
this SNR. The central bright belt (with larger electron density and longer
shocked time) show smaller ionization age than other regions.
All of the element abundance maps show no clear stratification. The central
barlike structure is not enhanced in any of the element abundance, supporting
its origin of shocked circumstellar material.
The Fe emission lines are not prominent among the whole remnant, and
its abundance are significantly deduced, indicating that the
reverse shock may have not propagated to the Fe-rich ejecta.
The O/Ne/Mg abundances show tight correlation with each other, and so do Si/S.
Such correlation suggests a common origin of nucleosynthesis
for O/Ne/Mg and Si/S respectively.
Based on relative abundances of O, Ne, Mg, Si and Fe to Si, we
suggest a progenitor mass of $25-30~M_{\odot}$ for G292.0+1.8.

\normalem

\begin{acknowledgements}
We acknowledge the use of data obtained by Chandra. The Chandra
Observatory Center is operated by the Smithsonian Astrophysical
Observatory for and on the behalf of NASA.
This project is supported by the National Natural Science Foundation of
China under Nos. 10903007 and 11273022.
\end{acknowledgements}

%
%

\begin{table*}
\caption[]{coherent coefficient for data points excluding extremum ones. The corresponding
value for Cas A (Yang et al. 2008) are given in parentheses. }
\label{coherence_CasA}
\begin{tabular}{ccccccc}
\noalign{\smallskip} \hline \hline \noalign{\smallskip}
Element&O&Ne&Mg&Si&S&Fe\\
\noalign{\smallskip} \hline \noalign{\smallskip}

O   &      $-$      &   0.64(0.33)   &   0.66(0.41)   &  0.58(0.33)  &   0.45(0.26)    &   0.43(0.13)   \\
Ne  &  0.64(0.33)   &      $-$       &   0.88(0.49)   &  0.47(0.19)  &   0.41(0.16)    &   0.03(0.25)   \\
Mg  &  0.66(0.41)   &   0.88(0.49)   &      $-$       &  0.52(0.48)  &   0.29(0.39)    &   0.24(0.24)   \\
Si  &  0.58(0.33)   &   0.47(0.19)   &   0.52(0.48)   &     $-$      &   0.70(0.86)    &   0.31(0.23)   \\
S   &  0.45(0.26)   &   0.41(0.16)   &   0.29(0.39)   &  0.70(0.86)  &       $-$       &   0.01(0.11)   \\
Fe  &  0.43(0.13)   &   0.03(0.25)   &   0.24(0.24)   &  0.31(0.23)  &   0.01(0.11)    &       $-$      \\

\noalign{\smallskip} \hline \noalign{\smallskip}
\end{tabular}
\end{table*}

\begin{table*}
\caption[]{Emission measure weighted average abundance and their ratios with respect to Si. }
\label{abundance_ratio}
\begin{tabular}{ccccccc}
\noalign{\smallskip} \hline \hline \noalign{\smallskip}
Element&O&Ne&Mg&Si&S&Fe\\
\noalign{\smallskip} \hline \noalign{\smallskip}

$Z$                     &      1.32      &   1.24   &   0.69   &  0.55  &   0.88    &   0.20   \\
$\sigma_{Z}$            &      1.03      &   0.41   &   0.09   &  0.07  &   0.38    &   0.01   \\
$Z/Z_{Si}$              &      2.40      &   2.24   &   1.25   &   $-$  &   1.59    &   0.37   \\
$\sigma_{Z/Z_{Si}}$     &      1.58      &   0.47   &   0.01   &   $-$  &   0.47    &   0.03   \\
$(Z/Z_{Si})^{\prime}$   &      3.29      &   2.54   &   1.30   &   $-$  &   1.53    &   0.35   \\



\noalign{\smallskip} \hline \noalign{\smallskip}
\end{tabular} \\
$Z$: emission measure weighted average abundance;           \\
$\sigma_{Z}$: weighted standard deviation for $Z$;          \\
$Z/Z_{Si}$: elemental abundance ratio with respect to Si;   \\
$\sigma_{Z/Z_{Si}}$: standard deviation for $Z/Z_{Si}$;     \\
$(Z/Z_{Si})^{\prime}$: emission measure weighted average abundance for the regions with O abundance greater than 1.0.           \\
\end{table*}

\end{document}